\RequirePackage{fix-cm}
\documentclass[smallextended]{svjour3}       % onecolumn (second format)
\smartqed  % flush right qed marks, e.g. at end of proof
\usepackage{graphicx}
\usepackage{amsmath,epsfig,amssymb,graphicx,caption}
\usepackage{subcaption}
\usepackage{setspace}
\begin{document}

%%%%%%%%%%%%%%%%%%%%%%%%%%%%%%%%%%%%%%%%%%%%%%%%%
%%% TITLE %%%%%%%%%%%%%%%%%%%%%%%%%%%%%%%%%%%%%%%
%%%%%%%%%%%%%%%%%%%%%%%%%%%%%%%%%%%%%%%%%%%%%%%%%
\title{Effective Capacity of Receive Antenna Selection MIMO-OSTBC Systems in Co-Channel Interference}
\author{Mohammad Lari}
\institute{Electrical and Computer Engineering Faculty \at
              Semnan University, Semnan, Iran \\
              Tel.: +9823-33383947\\
              \email{m\_lari@semnan.ac.ir}}
\date{Received: date / Accepted: date}
\maketitle

%%%%%%%%%%%%%%%%%%%%%%%%%%%%%%%%%%%%%%%%%%%%%%%%%
%%% ABSTRACT %%%%%%%%%%%%%%%%%%%%%%%%%%%%%%%%%%%%
%%%%%%%%%%%%%%%%%%%%%%%%%%%%%%%%%%%%%%%%%%%%%%%%%
\begin{abstract}
In this paper, delay constrained performance of a multiple-input multiple-output (MIMO) communication system in a dense environment with co-channel interference is investigated. We apply orthogonal space-time block coding (OSTBC) at the transmitter, and for alleviating the high complexity and cost of the MIMO system, receive antenna selection (RAS) scheme is employed in the downlink. Here, for simple and cheap mobile handsets, one antenna is chosen at the receiver in each utilization of the channel. Under these assumptions, a maximum constant arrival rate with the delay quality-of-service (QoS) guarantee in a wireless channel is extracted. We obtain a closed-form solution for the effective capacity of the MIMO-OSTBC channel with the RAS scheme in a quasi-static Rayleigh fading conditions and co-channel interference. After all, the numerical simulations are provided and verified the theoretical results.
\end{abstract}
%%%%%%%%%%%%%%%%%%%%%%%%%%%%%%%%%%%%%%%%%%%%%%%%%
%%% SECTION 1: INTRODUCTION %%%%%%%%%%%%%%%%%%%%%
%%%%%%%%%%%%%%%%%%%%%%%%%%%%%%%%%%%%%%%%%%%%%%%%%
\section{Introduction}\label{sec:1}
Delay sensitive applications such as audio/video conferencing and real-time control require low end-to-end delay \cite{Su}. In these applications, when the ultimate delay of a data packet exceeds a certain threshold, the packet is useless and has to be dropped. So, the delay violation probability is a suitable quality of service (QoS) performance metric here \cite{Niu}.

In wireless telecommunication systems and a proper channel condition, transmission with high data rate can be accomplished. In contrast, in a severe condition, the data rate decreases and probably becomes zero. Accordingly, data must tolerate longer stop at the transmitter buffer before transmission. Therefore, in the wireless channel, deterministic guarantee of total transmission delay for each data packet is not possible. However, with a genuine design, statistical delay constraint can be meet. 

In this respect, the interesting theory of effective capacity has been presented recently \cite{Negi}-\cite{Soret}. In contrast to Shannon capacity without any restrictions on complexity and delay, effective capacity is defined as the maximum constant arrival rate that a wireless channel can support in order to guarantee quality-of-service (QoS) requirements such as the statistical delay constraint \cite{Negi}-\cite{Soret}.

The theory of effective capacity in a general time correlated Rayleigh fading channel is introduced in \cite{Negi} and in the low signal to noise ratio (SNR) regime, a closed-form solution for the effective capacity is extracted. This concept with the statistical QoS constraint is completely discussed with a different viewpoint in \cite{Soret} for a generic traffic source over a time correlated Rayleigh fading channels. The authors assume a Gaussian distribution for the accumulated service rate of the channel, which leads to a solution for the effective capacity of the channels. Following these basic researches, the effective capacity has been conducted in various communication links such as rate and power adoption in single-input single-output (SISO) and multiple-input multiple-output (MIMO) systems \cite{ZhangSISO}-\cite{ZhangMIMO}, antenna selection (AS) MIMO systems \cite{Lari1}-\cite{Lari2} and  cooperative and cognitive radio channels in \cite{Lari3}-\cite{Aissa}. 

A MIMO wireless communication system is defined as a transmission link where the transmitter and the receiver are equipped with multiple antenna elements to increase channel capacity and data throughput. Digital communication using MIMO wireless link, has recently emerged as one of the most significant technical breakthroughs in modern communications such as IEEE 802.16 WiMax, IEEE 802.11 WiFi, cellular third and forth generation (3G/4G) systems \cite{lte}. Hence, for high data rate, multiple antenna structure is assumed here. Study of effective capacity in the MIMO channel is a new research topic which is discussed in receive/transmit antenna selection MIMO systems in \cite{Lari1}-\cite{Lari2}. In \cite{Lari1}, the effective capacity of a system in two different cases with single receive antenna selection (RAS) and single transmit antenna selection (TAS) schemes in quasi-static fading channel is investigated. When the channel state information (CSI) is not available at the transmitter, single RAS scheme is employed. On the other hand, with the available CSI at the transmitter, single TAS is executed. Moreover, an optimal power-control policy is applied to the selected antenna and the effective capacity of the MIMO system is derived. In \cite{Lari2}, a maximum constant arrival rate with the delay quality-of-service (QoS) guarantee is obtained in a MIMO orthogonal space time block coding (MIMO-OSTBC) system. The authors have assumed spatially correlated transmit antenna and simple receiver which employs single RAS technique. In both \cite{Lari1} and \cite{Lari2}, SNR maximization is considered as a criteria in the antenna selection procedure.

Regarding day by day increase of mobile users and request of high frequency reuse in the networks, interference mitigation is considered as one of the main challenge in modern communication systems \cite{lte}-\cite{duplex}. In this way, we discuss MIMO-OSTBC downlink scenario with co-channel interference here. This point is a reasonable assumption when the mobile station is near the cell edge or even when cognitive radio users are active in the same frequency \cite{Goldsmith}-\cite{Moustafa}. In the downlink, the transmitter usually have sufficient area, but, the receiver dimension is small. Therefore, OSTBC is employed at the transmitter side and single RAS technique is applied at the receiver. For this purpose, the receiver examines their antennas in each utilization of the channel, and select the antenna which maximizes the signal to noise plus interference ratio (SINR). Single RAS can simplify the receive RF front end \cite{LariBassam} while some performance of multiple antenna is preserved. Finally, the effective capacity of this MIMO-OSTBC system with single RAS scheme is investigated and approved by numerical simulations. The contribution of this paper can be summarized as:
\begin{itemize}
\item Closed-form calculation of the effective capacity in the MIMO-OSTBC downlink channel with the RAS scheme, when there are some co-channel interferers working simultaneously
\item Comparing the effective capacity in the MIMO-OSTBS systems with RAS and also MISO-OSTBC systems which shows the necessity of using RAS technique when high QoS is required (See Fig. 7)
\end{itemize}

This paper is organized as fallows: First, the system model is introduced in Section \ref{sec:2}. Section \ref{sec:3} provides a brief introduction to the effective capacity concept. The effective capacity of a MIMO-OSTBC system with the RAS scheme is explained in Section \ref{sec:4}. Finally, the simulation results are presented in Section \ref{sec:5}, and Section \ref{sec:6} concludes the paper.

\section{System Model}\label{sec:2}
\begin{figure}
\begin{center}
\includegraphics[draft=false,scale=0.4]{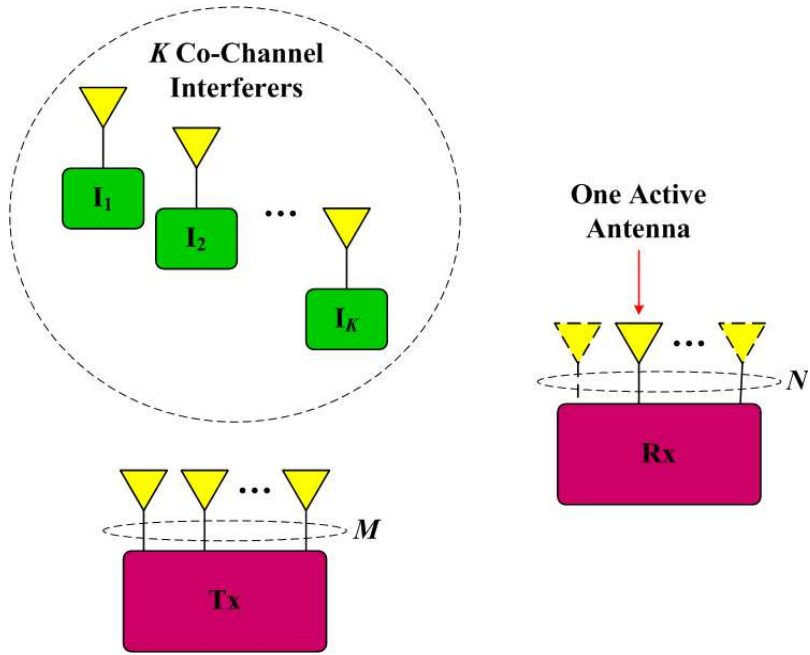}
\end{center}
\caption{System model.}
\label{fig:1}
\end{figure}

The downlink MIMO-OSTBC system with co-channel interferers is illustrated in Fig. \ref{fig:1}. The MIMO system has $M$ transmit and $N$ receive antennas and there are $K$ single antenna interferers which disturb our MIMO receiver in the same frequency. So, in addition to the additive white Gaussian noise, we encounter the co-channel interference at the MIMO receiver. 
%%%%%%%%%%%%%%%%%%%%%%%%%%%%%%%%%%%%%%%%%%%%%%%%%
%%% FIGURE 1 %%%%%%%%%%%%%%%%%%%%%%%%%%%%%%%%%%%%
%%%%%%%%%%%%%%%%%%%%%%%%%%%%%%%%%%%%%%%%%%%%%%%%%
In each utilization of the channel, $Q$ symbols are packed, encoded and construct the desired OSTBC. Next, this code transmits from $M$ antennas at $T_s$ time slots. Therefore, the code rate is considered as $R_c=Q/T_s$. We assume quasi-static Rayleigh flat fading channel in both MIMO link and interferer channel. According to this model, the received symbols by $N$ receive antennas at $T_s$ time slots is given by
%%%%%%%%%%%%%%%%%%%%%%%%%%%%%%%%%%%%%%%%%%%%%%%%%
%%% EQUATION - 1 %%%%%%%%%%%%%%%%%%%%%%%%%%%%%%%%
%%%%%%%%%%%%%%%%%%%%%%%%%%%%%%%%%%%%%%%%%%%%%%%%%
\begin{equation}\label{eq:1}
\mathbf{R}=\mathbf{S}\times\mathbf{H}+\mathbf{C}\times\mathbf{G}+\mathbf{W}
\end{equation}
where $\mathbf{S}\in\mathcal{C}^{T_s\times M}$ represents the OSTBC transmitted symbols and $\mathbf{C}\in\mathcal{C}^{T_s\times K}$ contains transmitted symbols from $K$ co-channel interferers during $T_s$ time slots. Similarly, $\mathbf{H}\in\mathcal{C}^{M\times N}$ determine the quasi-static flat fading channels between $M$ transmit and $N$ receive antennas, $\mathbf{G}\in\mathcal{C}^{K\times N}$ is the channel coefficients between $K$ interferers and MIMO receiver and $W\in\mathcal{C}^{T_s\times N}$ represents additive white Gaussian noise components at $N$ antennas and $T_s$ time slots. Note that, $\mathcal{C}^{A\times B}$ displays $A\times B$ complex matrix set.

We assume single RAS scheme in the receiver. Therefore, we encounter multiple-input single-output (MISO) system for more simplicity. Regarding (\ref{eq:1}), the received symbols by the selected antenna at $T_s$ time slots is simplified to
%%%%%%%%%%%%%%%%%%%%%%%%%%%%%%%%%%%%%%%%%%%%%%%%%
%%% EQUATION - 2 %%%%%%%%%%%%%%%%%%%%%%%%%%%%%%%%
%%%%%%%%%%%%%%%%%%%%%%%%%%%%%%%%%%%%%%%%%%%%%%%%%
\begin{equation}\label{eq:2}
\mathbf{r}=\mathbf{S}\times\mathbf{h}+\mathbf{C}\times\mathbf{g}+\mathbf{w}
\end{equation}
where $\mathbf{h}\in\mathcal{C}^{M\times 1}$ depicts MISO channel coefficients ans $\mathbf{g}\in\mathcal{C}^{K\times 1}$ shows interferers channel components.We will show the elements of $\mathbf{h}$ and $\mathbf{g}$ by $\{h_m\}_{m=1,2,...,M}$ and $\{g_k\}_{k=1,2,...,K}$ respectively.

\begin{figure}
\begin{center}
\includegraphics[draft=false,width=\linewidth]{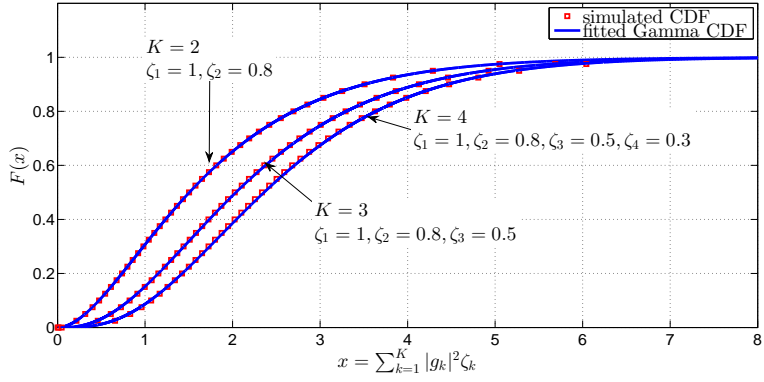}
\end{center}
\caption{Simulated and fitted CDF of $\sum_{k=1}^{K}|g_k|^2\zeta_k$.}
\label{fig:2}
\end{figure}

In the MISO-OSTBC channels without any distortion, received SNR is expressed as \cite{Shin}
%%%%%%%%%%%%%%%%%%%%%%%%%%%%%%%%%%%%%%%%%%%%%%%%%
%%% EQUATION - 3 %%%%%%%%%%%%%%%%%%%%%%%%%%%%%%%%
%%%%%%%%%%%%%%%%%%%%%%%%%%%%%%%%%%%%%%%%%%%%%%%%%
\begin{equation}\label{eq:3}
\gamma_{\mathrm{STBC}} = \frac{\gamma_0}{MR_c}\mathbf{h}^{\dagger}\times \mathbf{h}= \frac{\gamma_0}{MR_c}\sum_{m=1}^{M}|h_m|^2
\end{equation}
where $\gamma_0$ is the average SNR of the link. However, when some interferers are available, the transmitted signals by the interferer users are added to the noise and similar to (\ref{eq:3}), we can write signal to interference plus noise ratio (SINR) as
%%%%%%%%%%%%%%%%%%%%%%%%%%%%%%%%%%%%%%%%%%%%%%%%%
%%% EQUATION - 4 %%%%%%%%%%%%%%%%%%%%%%%%%%%%%%%%
%%%%%%%%%%%%%%%%%%%%%%%%%%%%%%%%%%%%%%%%%%%%%%%%%
\begin{equation}\label{eq:4}
\gamma_{\mathrm{STBC}} = \frac{\gamma_0}{MR_c}\frac{\mathbf{h}^{\dagger}\times \mathbf{h}}{1+\sum_{k=1}^{K}|g_k|^2\mathcal{P}_k/\mathcal{P}_w}=\frac{\gamma_0}{MR_c}\frac{\sum_{m=1}^{M}|h_m|^2}{1+\sum_{k=1}^{K}|g_k|^2\zeta_k}.
\end{equation}
Here, $\mathcal{P}_k$ represents the received power by the $k$-th interferers at the MISO receiver, $\mathcal{P}_w$ is the white noise power and $\zeta_k=\mathcal{P}_k/\mathcal{P}_w$ shows the power of
$k$-th interferer to noise ratio. Note that, in a dense environment or when we have very close interferers, we can assume $\sum_{k=1}^{K}|g_k|^2\zeta_k\gg1$, so, the SINR in (\ref{eq:4}) is approximated as 
%%%%%%%%%%%%%%%%%%%%%%%%%%%%%%%%%%%%%%%%%%%%%%%%%
%%% EQUATION - 5 %%%%%%%%%%%%%%%%%%%%%%%%%%%%%%%%
%%%%%%%%%%%%%%%%%%%%%%%%%%%%%%%%%%%%%%%%%%%%%%%%%
\begin{equation}\label{eq:5}
\gamma_{\mathrm{STBC}} = \frac{\gamma_0}{MR_c}\frac{\sum_{m=1}^{M}|h_m|^2}{\sum_{k=1}^{K}|g_k|^2\zeta_k}.
\end{equation}
If all interferers have nearly the same received power, we have $\zeta_k\approx\zeta$ and then, $\sum_{k=1}^{K}|g_k|^2\zeta_k=\zeta\sum_{k=1}^{K}|g_k|^2$ has Gamma distribution. However, when $\zeta_k$ is not the same for $K$ interferers, $\sum_{k=1}^{K}|g_k|^2\zeta_k$ is very tightly approximated as a Gamma distributed random number. This claim is demonstrated for the cumulative distribution function (CDF) of $K=2,3,4$ interferers in Fig. \ref{fig:2}. The values of $\zeta_k$ for different $K$, have displayed in Fig. \ref{fig:2}. Now, for more simplicity, we assume $\zeta_k=\zeta$ in the proceeding. 
%%%%%%%%%%%%%%%%%%%%%%%%%%%%%%%%%%%%%%%%%%%%%%%%%
%%% FIGURE 2 %%%%%%%%%%%%%%%%%%%%%%%%%%%%%%%%%%%%
%%%%%%%%%%%%%%%%%%%%%%%%%%%%%%%%%%%%%%%%%%%%%%%%%

Since $\sum_{m=1}^{M}|h_m|^2$ and $\sum_{k=1}^{K}|g_k|^2$ have Gamma distribution ($\sum_{m=1}^{M}|h_m|^2\sim \mathrm{Gamma}(\alpha=M,\beta=1)$, $\sum_{k=1}^{K}|g_k|^2\sim \mathrm{Gamma}(\alpha=K,\beta=1)$), then, $\Omega=\sum_{m=1}^{M}|h_m|^2/\sum_{k=1}^{K}|g_k|^2$ has Beta-prime distribution ($\Omega\sim\mathrm{Beta-prime}(\phi_1=M,\phi_2=K)$) \cite{Yao} with the probability density function (PDF) of
%%%%%%%%%%%%%%%%%%%%%%%%%%%%%%%%%%%%%%%%%%%%%%%%%
%%% EQUATION - 6 %%%%%%%%%%%%%%%%%%%%%%%%%%%%%%%%
%%%%%%%%%%%%%%%%%%%%%%%%%%%%%%%%%%%%%%%%%%%%%%%%%
\begin{equation}\label{eq:6}
f_{\Omega}(x)=\frac{x^{M-1}(1+x)^{-M-K}}{\mathcal{B}(M,K)},
\end{equation}
and CDF of
%%%%%%%%%%%%%%%%%%%%%%%%%%%%%%%%%%%%%%%%%%%%%%%%%
%%% EQUATION - 7 %%%%%%%%%%%%%%%%%%%%%%%%%%%%%%%%
%%%%%%%%%%%%%%%%%%%%%%%%%%%%%%%%%%%%%%%%%%%%%%%%%
\begin{equation}\label{eq:7}
F_{\Omega}(x)=\int_0^x f_\Omega(\lambda)d\lambda
\end{equation}
where $\mathcal{B}(.,.)$ stands for the Beta function \cite{Ryzhik}. Next, we summarize the SINR as $\gamma_{\mathrm{STBC}} = \frac{\xi_0}{MR_c}\Omega$ where $\gamma_0/\zeta$ is replaced by $\xi_0$, and $\xi_0$ denotes the signal to interference ratio as well. To obtain the solution of (\ref{eq:7}), we can use \cite[eq. 2.111-2]{Ryzhik} and we get
%%%%%%%%%%%%%%%%%%%%%%%%%%%%%%%%%%%%%%%%%%%%%%%%%
%%% EQUATION - 8 %%%%%%%%%%%%%%%%%%%%%%%%%%%%%%%%
%%%%%%%%%%%%%%%%%%%%%%%%%%%%%%%%%%%%%%%%%%%%%%%%%
\begin{align}\label{eq:8}
&F_{\Omega}(x)=\frac{(-1)^{M}(M-1)!}{\mathcal{B}(M,K)\prod_{m=1}^{M}(m-M-K)}\times\nonumber\\
&\left\lbrace\sum_{k=1}^{M+K-1}{M+K-1 \choose k}\frac{x^k}{(1+x)^{M+K-1}}+\sum_{m=1}^{M-1}\frac{(-1)^{m+1}\prod_{\mathfrak{m}=1}^{m}(\mathfrak{m}-M-K)}{m!}\frac{x^{m}}{(1+x)^{M+K-1}}\right\rbrace\nonumber\\
&\qquad=\frac{(-1)^{M}(M-1)!}{\mathcal{B}(M,K)\prod_{m=1}^{M}(m-M-K)}\sum_{k=1}^{M+K-1}p_k\frac{x^k}{(1+x)^{M+K-1}},
\end{align}
where
%%%%%%%%%%%%%%%%%%%%%%%%%%%%%%%%%%%%%%%%%%%%%%%%%
%%% EQUATION - 9 %%%%%%%%%%%%%%%%%%%%%%%%%%%%%%%%
%%%%%%%%%%%%%%%%%%%%%%%%%%%%%%%%%%%%%%%%%%%%%%%%%
\begin{equation}\label{eq:9}
p_k=
\begin{cases}
{M+K-1 \choose k}+\frac{(-1)^{k+1}\prod_{\mathfrak{k}=1}^{k}(\mathfrak{k}-M-K)}{k!}&,1\le k\le M-1\\
{M+K-1 \choose k}&,M\le k\le M+K-1
\end{cases}.
\end{equation}
\section{The Effective Capacity}\label{sec:3}
Despite the time-varying nature of wireless channels, network service providers must guarantee a specific QoS to satisfy their customers who need real-time multimedia communication. In \cite{Negi}, Wu and Negi defined the effective capacity as the maximum constant arrival rate that a given service process can support in order to guarantee a QoS requirement specified by the QoS exponent $\theta$. For a dynamic queuing system with stationary and ergodic arrival and service processes, under sufficient conditions, the queue length process $Q(t)$ converges in distribution to a random variable $Q(\infty)$ such that \cite{Negi}
%%%%%%%%%%%%%%%%%%%%%%%%%%%%%%%%%%%%%%%%%%%%%%%%%
%%% EQUATION 10 %%%%%%%%%%%%%%%%%%%%%%%%%%%%%%%%%
%%%%%%%%%%%%%%%%%%%%%%%%%%%%%%%%%%%%%%%%%%%%%%%%%
\begin{equation}\label{eq:10}
-\lim_{q\to \infty}\frac{\ln\left(\textrm{Pr}\{Q(\infty)>q\}\right)}{q}=\theta,
\end{equation}
where $q$ determines a certain threshold, and therefore, we have the following approximation:
%%%%%%%%%%%%%%%%%%%%%%%%%%%%%%%%%%%%%%%%%%%%%%%%%
%%% EQUATION 11 %%%%%%%%%%%%%%%%%%%%%%%%%%%%%%%%%
%%%%%%%%%%%%%%%%%%%%%%%%%%%%%%%%%%%%%%%%%%%%%%%%%
\begin{equation}\label{eq:11}
\textrm{Pr}\{Q(\infty)>q\}\approx e^{-\theta q}
\end{equation}
for a large $q$. For a small $q$ the following approximation is shown to be more accurate \cite{Negi}:
%%%%%%%%%%%%%%%%%%%%%%%%%%%%%%%%%%%%%%%%%%%%%%%%%
%%% EQUATION 12 %%%%%%%%%%%%%%%%%%%%%%%%%%%%%%%%%
%%%%%%%%%%%%%%%%%%%%%%%%%%%%%%%%%%%%%%%%%%%%%%%%%
\begin{equation}\label{eq:12}
\textrm{Pr}\{Q(\infty)>q\}\approx \varepsilon e^{-\theta q},
\end{equation}
where $\varepsilon$ represents the non-empty buffer probability. In addition, for the delay experienced by a packet as the main QoS metric, we have a similar probability function as \cite{Negi}
%%%%%%%%%%%%%%%%%%%%%%%%%%%%%%%%%%%%%%%%%%%%%%%%%
%%% EQUATION 13 %%%%%%%%%%%%%%%%%%%%%%%%%%%%%%%%%
%%%%%%%%%%%%%%%%%%%%%%%%%%%%%%%%%%%%%%%%%%%%%%%%%
\begin{equation}\label{eq:13}
\textrm{Pr}\{D>d\}\approx\varepsilon e^{-\theta \delta d},
\end{equation}
where $D$ indicates the tolerated delay, $d$ is a delay-bound, and $\delta$ is jointly determined by both arrival and service processes. The statistical delay constraint in (\ref{eq:13}) represents the QoS which has to be guaranteed for the delay sensitive traffic sources. It is apparent that the QoS exponent $\theta$ has an important role here. Larger $\theta$ corresponds to more strict QoS constraint, while smaller $\theta$ implies looser QoS requirements.

Effective capacity provides the maximum constant arrival rate that can be supported by the time-varying wireless channel under the statistical delay constraint (\ref{eq:13}). Since the average arrival rate is equal to the average departure rate when the buffer is in a steady-state \cite{Chang}, the effective capacity can be viewed as the maximum throughput in the presence of such a constraint. The effective capacity is defined in \cite[eq. 12]{Negi} and \cite[eq. 6]{Soret} in the correlated channels. However in the uncorrelated case it reduces to
%%%%%%%%%%%%%%%%%%%%%%%%%%%%%%%%%%%%%%%%%%%%%%%%%
%%% EQUATION 14 %%%%%%%%%%%%%%%%%%%%%%%%%%%%%%%%%
%%%%%%%%%%%%%%%%%%%%%%%%%%%%%%%%%%%%%%%%%%%%%%%%%
\begin{equation}\label{eq:14}
E_C(\theta)=-\frac{1}{\theta}\ln\left(\textsf{E}\left\{e^{-\theta R}\right\}\right)
\end{equation}
where $R$ is the time-varying rate of the channel and $\textsf{E}\{.\}$ denotes the expectation. For a specific application with a given statistical delay requirement, the QoS exponent $\theta$ can be determined from (\ref{eq:13}). Then, the maximum constant arrival rate of the sources that a wireless channel can support in order to guarantee the given QoS, is determined from (\ref{eq:14}). $R$ and the effective capacity $E_C(\theta)$ are further discussed in the next section.

\section{Effective Capacity of MIMO-OSTBC with Co-Channel Interference}\label{sec:4}
The physical layer of downlink MIMO-OSTBC system is introduced in Section \ref{sec:2} where for obtaining simple and cheap handsets, we assumed single RAS technique at the receiver. In addition, single RAS can eliminate practical issues such as mutual coupling and spatial correlation between antennas \cite{Catreux}. In our assumed case, in each utilization of the channel, one of the $N$ receive antenna is selected. This antenna has maximum SINR among the all $N$ antennas as well. Here, $\{\gamma_{\mathrm{STBC}}^1=\frac{\xi_0}{MR_c}\Omega^1,\gamma_{\mathrm{STBC}}^2=\frac{\xi_0}{MR_c}\Omega^2,...,\gamma_{\mathrm{STBC}}^N=\frac{\xi_0}{MR_c}\Omega^N\}$ denotes the SINR at $N$ different receive antennas, and $\{\gamma_{\mathrm{STBC}}^{(1)}=\frac{\xi_0}{MR_c}\Omega^{(1)},\gamma_{\mathrm{STBC}}^{(2)}=\frac{\xi_0}{MR_c}\Omega^{(2)},...,\gamma_{\mathrm{STBC}}^{(N)}=\frac{\xi_0}{MR_c}\Omega^{(N)}\}$ represents these random numbers in an ordered form where $\{\gamma_{\mathrm{STBC}}^{(1)}\leq\gamma_{\mathrm{STBC}}^{(2)}\leq...\leq\gamma_{\mathrm{STBC}}^{(N)}\}$. Therefore, the channel capacity in this scenario is written as
%%%%%%%%%%%%%%%%%%%%%%%%%%%%%%%%%%%%%%%%%%%%%%%%%
%%% EQUATION 15 %%%%%%%%%%%%%%%%%%%%%%%%%%%%%%%%%
%%%%%%%%%%%%%%%%%%%%%%%%%%%%%%%%%%%%%%%%%%%%%%%%%
\begin{equation}\label{eq:15}
R=BT_fR_c\log_2\left(1+\gamma_{\mathrm{STBC}}^{(N)}\right)=BT_fR_c\log_2\left(1+\frac{\xi_0}{MR_c}\Omega^{(N)}\right)
\end{equation}
where $B$ is the total spectral bandwidth, $T_f$ is the frame duration, and $\Omega^{(N)}$  is the maximum value in the set of $\{\Omega^1,\Omega^2,...,\Omega^N\}$. By inserting (\ref{eq:15}) into (\ref{eq:14}), we obtain the effective capacity as 
%%%%%%%%%%%%%%%%%%%%%%%%%%%%%%%%%%%%%%%%%%%%%%%%%
%%% EQUATION 16 %%%%%%%%%%%%%%%%%%%%%%%%%%%%%%%%%
%%%%%%%%%%%%%%%%%%%%%%%%%%%%%%%%%%%%%%%%%%%%%%%%%
\begin{equation}\label{eq:16}
E_C(\theta)=-\frac{1}{\theta}\ln\left(\textsf{E}\left\{\left(1+\frac{\xi_0}{MR_c}\Omega^{(N)}\right)^{-\breve{\theta}}\right\}\right)
\end{equation}
and, $\breve{\theta}=\theta BT_fR_c/\ln2$.

Since $\Omega^n, n=1,2,...,N$ has Beta-prime distribution $(\Omega^n\sim\mathrm{Beta-prime}(\phi_1=M,\phi_2=K))$, the PDF of $\Omega^{(N)}$ can be found as \cite{David}
%%%%%%%%%%%%%%%%%%%%%%%%%%%%%%%%%%%%%%%%%%%%%%%%%
%%% EQUATION 17 %%%%%%%%%%%%%%%%%%%%%%%%%%%%%%%%%
%%%%%%%%%%%%%%%%%%%%%%%%%%%%%%%%%%%%%%%%%%%%%%%%%
\begin{equation}\label{eq:17}
f_{\Omega^{(N)}}(x)=Nf_{\Omega}(x)\left[F_{\Omega}(x)\right]^{N-1}
\end{equation}
where $f_{\Omega}(x)$ and $F_{\Omega}(x)$ were determined in (\ref{eq:6}) and (\ref{eq:8}) respectively.
Now, by employing multinomial theorem, we get
%%%%%%%%%%%%%%%%%%%%%%%%%%%%%%%%%%%%%%%%%%%%%%%%%
%%% EQUATION 18 %%%%%%%%%%%%%%%%%%%%%%%%%%%%%%%%%
%%%%%%%%%%%%%%%%%%%%%%%%%%%%%%%%%%%%%%%%%%%%%%%%%
\begin{align}\label{eq:18}
&\left[F_{\Omega}(x)\right]^{N-1}=\nonumber\\
&\left[\frac{(-1)^{M}(M-1)!}{\mathcal{B}(M,K)\prod_{m=1}^{M}(m-M-K)}\right]^{N-1}\sum_{k=N-1}^{(N-1)(M+K-1)}q_{k-(N-1)}\frac{x^k}{(1+x)^{(N-1)(M+K-1)}},
\end{align}
where $q_{k-(N-1)}$ represents the $k-(N-1)$-th coefficient from the multinomial expansion of 
\begin{equation*}
\left[\sum_{k=1}^{M+K-1}p_k\frac{x^k}{(1+x)^{M+K-1}}\right]^{N-1}.
\end{equation*}
For simple calculation of $q_k$, we define $\vec{\textbf{q}}^{(1)}=\left[p_1, p_2,...,p_{M+K-1}\right]^T$ where $T$ shows the transpose operator and $\vec{\textbf{q}}^{(n)}=\vec{\textbf{q}}^{(n-1)}\otimes\vec{\textbf{q}}^{(1)}$ and $\otimes$ represents the discrete convolution function. Here, $q_k$ indicates the $k$-th element of $\vec{\textbf{q}}^{(N-1)}$. According to (\ref{eq:17}) and (\ref{eq:18}), $f_{\Omega^{(N)}}(x)$ is simplified to
%%%%%%%%%%%%%%%%%%%%%%%%%%%%%%%%%%%%%%%%%%%%%%%%%
%%% EQUATION 19 %%%%%%%%%%%%%%%%%%%%%%%%%%%%%%%%%
%%%%%%%%%%%%%%%%%%%%%%%%%%%%%%%%%%%%%%%%%%%%%%%%%
\begin{align}\label{eq:19}
&f_{\Omega^{(N)}}(x)=\nonumber\\
&\frac{N}{\mathcal{B}(M,K)}\left[\frac{(-1)^{M}(M-1)!}{\mathcal{B}(M,K)\prod_{m=1}^{M}(m-M-K)}\right]^{N-1}\sum_{k=N-1}^{(N-1)(M+K-1)}q_{k-(N-1)}\frac{x^{M+k-1}}{(1+x)^{N(M+K-1)+1}}
\end{align}
and we have
%%%%%%%%%%%%%%%%%%%%%%%%%%%%%%%%%%%%%%%%%%%%%%%%%
%%% EQUATION 20 %%%%%%%%%%%%%%%%%%%%%%%%%%%%%%%%%
%%%%%%%%%%%%%%%%%%%%%%%%%%%%%%%%%%%%%%%%%%%%%%%%%
\begin{align}\label{eq:20}
&\textsf{E}\left\{\left(1+\frac{\xi_0}{MR_c}\Omega^{(N)}\right)^{-\breve{\theta}}\right\}=\int_0^{\infty}\left(1+\frac{\xi_0}{MR_c}x\right)^{-\breve{\theta}}f_{\Omega^{(N)}}(x)dx\nonumber\\&=\frac{N}{\mathcal{B}(M,K)}\left[\frac{(-1)^{M}(M-1)!}{\mathcal{B}(M,K)\prod_{m=1}^{M}(m-M-K)}\right]^{N-1}\nonumber\\&\times\sum_{k=N-1}^{(N-1)(M+K-1)}\left[q_{k-(N-1)}\mathcal{B}(M+k,N(M+K-1)+1+\breve{\theta}-(M+k))\right.\nonumber\\&\left.\qquad\qquad\qquad\qquad\times{}_2F_1(\breve{\theta},M+k;N(M+K-1)+1+\breve{\theta};1-\xi_0/MR_c)\right]
\end{align}
 where ${}_2F_1(.,.;.;.)$ shows Gaussian hypergeometric function \cite{Ryzhik}. Note that, a closed-form solution for  (\ref{eq:20}) can be found by using the integral in \cite[eq. 3.197-5]{Ryzhik}. It is visible that (\ref{eq:20}) contains a finite sum of two special math functions and can be evaluated by commercial numeric software. By applying this expression into (\ref{eq:16}), eventually, the effective capacity of MIMO-OSTBC with RAS scheme is attained when $K$ co-channel interferers are alive. 

\section{Simulation Results}\label{sec:5}
Here we assume $B=100\textrm{KHz}$, $T_f=1\textrm{msec}$, $R_c=1$ and for satisfactory results, Mont-Carlo simulation is repeated $1,000,000$ times in each step. Moreover, for more simple drawing, $\overline{E_C}(\theta)=E_C(\theta)/(BT_f)$ is plotted in the followings.

\begin{figure}
\begin{center}
\includegraphics[draft=false,width=\linewidth]{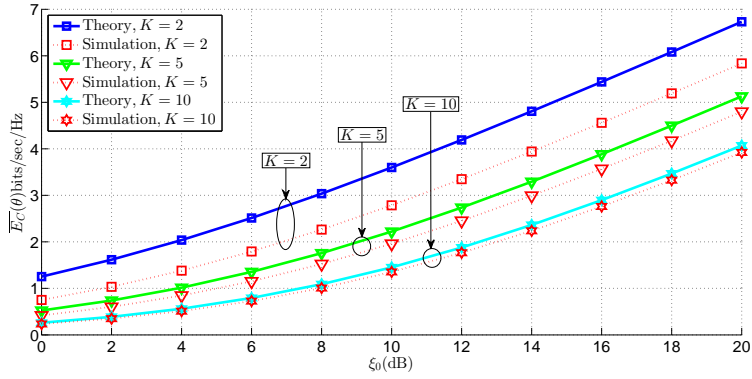}
\end{center}
\caption{Normalized effective capacity in a $2\times 4$ MIMO-OSTBC with RAS scheme versus the signal to interference ratio $\xi_0$, $\zeta=0\textrm{dB}$ and different number of interferers.}
\label{fig:3}
\end{figure}

\begin{figure}
\begin{center}
\includegraphics[draft=false,width=\linewidth]{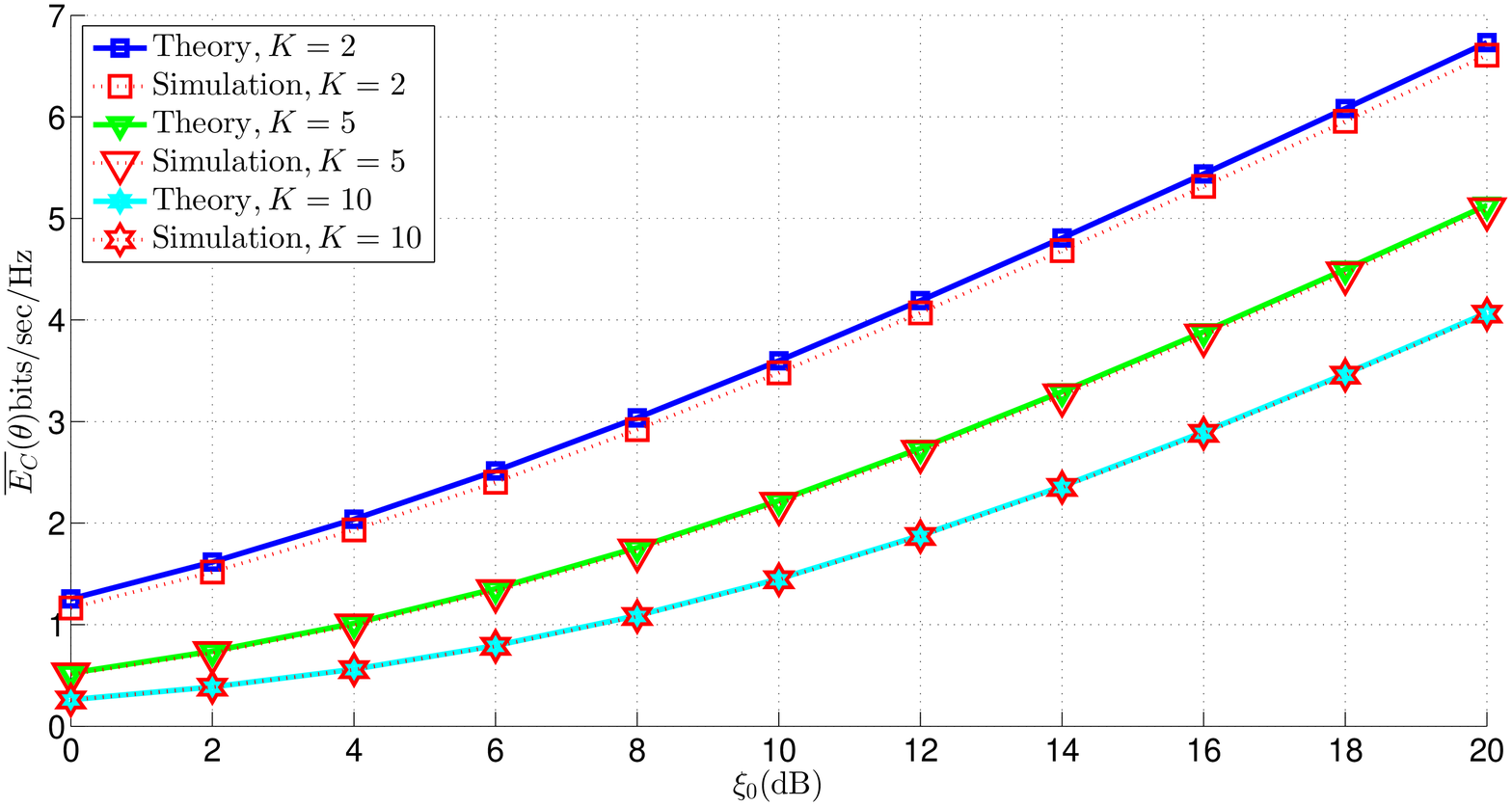}
\end{center}
\caption{Normalized effective capacity in a $2\times 4$ MIMO-OSTBC with RAS scheme versus the signal to interference ratio $\xi_0$, $\zeta=10\textrm{dB}$ and different number of interferers.}
\label{fig:4}
\end{figure}

\begin{figure}
\begin{center}
\includegraphics[draft=false,width=\linewidth]{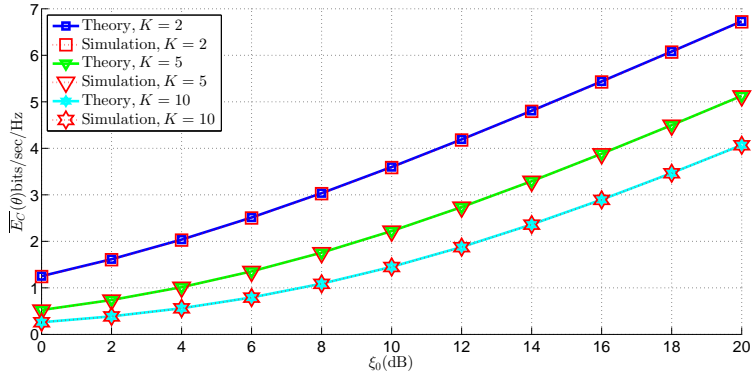}
\end{center}
\caption{Normalized effective capacity in a $2\times 4$ MIMO-OSTBC with RAS scheme versus the signal to interference ratio $\xi_0$, $\zeta=20\textrm{dB}$ and different number of interferers.}
\label{fig:5}
\end{figure}

The normalized effective capacity is plotted in Fig. \ref{fig:3}, \ref{fig:4} and Fig. \ref{fig:5} versus the signal to interference ratio $\xi_0$ in a MIMO-OSTBC systems and single RAS scheme. Also we assume $M=2$ and $N=4$, so that, one antenna is chosen in each utilization of the channel. In addition, the interference to noise ratio is equal to $\zeta=0\textrm{dB}$, $\zeta=10\textrm{dB}$ and $\zeta=20\textrm{dB}$ in Fig. \ref{fig:3}, \ref{fig:4} and Fig. \ref{fig:5} respectively and the simulation is executed for different number of co-channel interferers $K=2,5,10$ and $\theta=0.01$.

Tight agreement between theory and simulation, specially at high interference to noise ratio $\zeta$, or large number of interferers $K$ is clear. Therefore, the approximation of SINR in (\ref{eq:5}) is quit precise. Furthermore, we observe that increasing $\xi_0$ leads to increase of the effective capacity as might be expected. Again, increment of interferers causes some grade of decrease in the amount of effective capacity. hence, large number of interferers have to be avoid for obtaining suitable capacity with the QoS guarantee. 
%%%%%%%%%%%%%%%%%%%%%%%%%%%%%%%%%%%%%%%%%%%%%%%%%
%%% FIGURE 3 %%%%%%%%%%%%%%%%%%%%%%%%%%%%%%%%%%%%
%%%%%%%%%%%%%%%%%%%%%%%%%%%%%%%%%%%%%%%%%%%%%%%%%
%%%%%%%%%%%%%%%%%%%%%%%%%%%%%%%%%%%%%%%%%%%%%%%%%
%%% FIGURE 4 %%%%%%%%%%%%%%%%%%%%%%%%%%%%%%%%%%%%
%%%%%%%%%%%%%%%%%%%%%%%%%%%%%%%%%%%%%%%%%%%%%%%%%
%%%%%%%%%%%%%%%%%%%%%%%%%%%%%%%%%%%%%%%%%%%%%%%%%
%%% FIGURE 5 %%%%%%%%%%%%%%%%%%%%%%%%%%%%%%%%%%%%
%%%%%%%%%%%%%%%%%%%%%%%%%%%%%%%%%%%%%%%%%%%%%%%%%

\begin{figure}
  \begin{subfigure}[b]{\linewidth}
	\centering
	\includegraphics[draft=false,width=\linewidth]{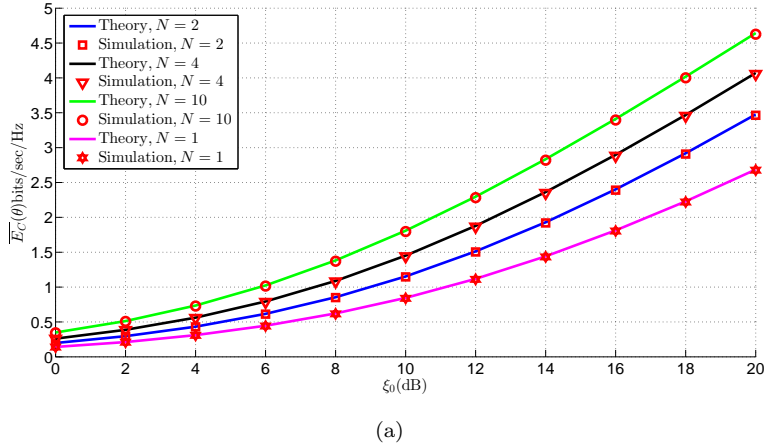}
	\caption{}
  \end{subfigure}\\
  \begin{subfigure}[b]{\linewidth}
  	\centering
    \includegraphics[draft=false,width=\linewidth]{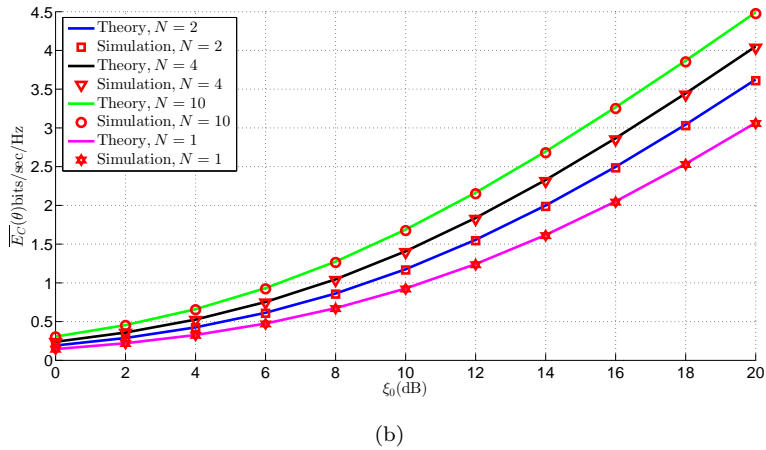}
    \caption{}
  \end{subfigure}
  \caption{Normalized effective capacity in a MIMO-OSTBC with RAS scheme versus the signal to interference ratio $\xi_0$, $\zeta=10\textrm{dB}$ and $K=10$ when (a) $M=2$ (b) $M=4$.}
  \label{fig:6}
\end{figure}

In Fig. \ref{fig:6} the normalized effective capacity is plotted versus the signal to interference ratio $\xi_0$ and $\zeta=10\textrm{dB}$, $K=10$ and $M=2$ in Fig. \ref{fig:6} (a) and $M=4$ in Fig. \ref{fig:6} (b). Fine agreement between theory and simulation results are clear. In addition, performance improvement is observed when the number of receive antenna is increased and one of them is chosen for signal reception. For more details, the normalized effective capacity of a MISO system without receive antenna selection is also plotted. Therefore, high performance is obtained with the low complexity of MISO system when RAS technique is applied at our MIMO systems.
%%%%%%%%%%%%%%%%%%%%%%%%%%%%%%%%%%%%%%%%%%%%%%%%%
%%% FIGURE 6 %%%%%%%%%%%%%%%%%%%%%%%%%%%%%%%%%%%%
%%%%%%%%%%%%%%%%%%%%%%%%%%%%%%%%%%%%%%%%%%%%%%%%%

\begin{figure}
\begin{center}
\includegraphics[draft=false,width=\linewidth]{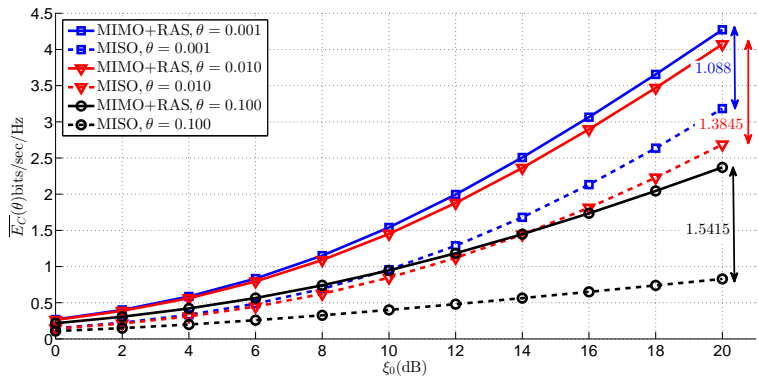}
\end{center}
\caption{Normalized effective capacity in a $2\times 4$ MIMO-OSTBC with RAS scheme and a $2\times 1$ MISO-OSTBC system versus the signal to interference ratio $\xi_0$ when $\zeta=10\textrm{dB}$ and $K=10$.}
\label{fig:7}
\end{figure}

The advantage of RAS scheme in the MIMO-OSTBC compared to the MISO-OSTBC systems is depicted in Fig. \ref{fig:7}. We assume a $2\times 4$ MIMO-OSTBC with RAS scheme and a $2\times 1$ MISO-OSTBC systems and plot the normalized effective capacity for three different QoS requirement. Also, $\zeta=10\textrm{dB}$ and $K=10$. It can be seen that the RAS is more efficient when high QoS (high value of $\theta$) at moderate to high SNR is requested. This point is specified in the figure where we have $1.5415 \textrm{ bits/sec/Hz}$ effective capacity gain at $\theta=0.1$, $1.3845 \textrm{ bits/sec/Hz}$ effective capacity gain at $\theta=0.01$ and $1.0880 \textrm{ bits/sec/Hz}$ effective capacity gain at $\theta=0.001$ respectively at $\xi_0=20\textrm{dB}$. Consequently, RAS technique can improve the performance, specially at high QoS request, and keep the complexity as low as a MISO system.

%%%%%%%%%%%%%%%%%%%%%%%%%%%%%%%%%%%%%%%%%%%%%%%%%
%%% FIGURE 7 %%%%%%%%%%%%%%%%%%%%%%%%%%%%%%%%%%%%
%%%%%%%%%%%%%%%%%%%%%%%%%%%%%%%%%%%%%%%%%%%%%%%%%

\begin{figure}
\begin{center}
\includegraphics[draft=false,width=\linewidth]{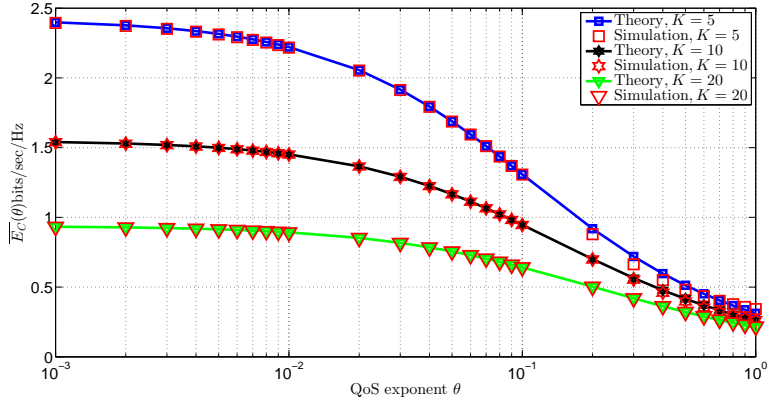}
\end{center}
\caption{Normalized effective capacity in a $2\times 4$ MIMO-OSTBC with RAS scheme versus the QoS exponent $\theta$ and $\xi_0=10\textrm{dB}$, $\zeta=10\textrm{dB}$ with $K=5$, $K=10$ or $K=20$ co-channel interferers.}
\label{fig:8}
\end{figure}

Finally, we plot the normalized effective capacity versus $\theta$ in Fig. \ref{fig:8}. We assume $M=2$, $N=4$, $\xi_0=10\textrm{dB}$ and $\zeta=10\textrm{dB}$ when we have $K=5$, $K=10$ or $K=20$ co-channel interferers. Similar to the previous figures, the simulation results follow the theory tightly. The normalized effective capacity decreases as the QoS exponent $\theta$ increases. Also, we observe that, by increase of co-channel interferers, the effective capacity decreases dramatically. Therefore,  using some interference cancellation techniques is necessary when good performance is expected. Note that, all curves converge to a single point at large QoS exponent $\theta$. 
%%%%%%%%%%%%%%%%%%%%%%%%%%%%%%%%%%%%%%%%%%%%%%%%%
%%% FIGURE 8 %%%%%%%%%%%%%%%%%%%%%%%%%%%%%%%%%%%%
%%%%%%%%%%%%%%%%%%%%%%%%%%%%%%%%%%%%%%%%%%%%%%%%%

%%%%%%%%%%%%%%%%%%%%%%%%%%%%%%%%%%%%%%%%%%%%%%%%%
%%% SECTION 6 %%%%%%%%%%%%%%%%%%%%%%%%%%%%%%%%%%%
%%%%%%%%%%%%%%%%%%%%%%%%%%%%%%%%%%%%%%%%%%%%%%%%%
\section{Conclusion}\label{sec:6}
Effective capacity is an interesting topic, so, study of this subject in a practical MIMO system with RAS  scheme under co-channel interference circumstance is considered here. To obtain simple and cheap MIMO receiver in downlink, we adopt RAS technique and a closed-form solution for the maximum constant arrival rate with the QoS guarantee is extracted. The solution is tightly accurate when large number of co-channel interferers or some strong co-channel interferers are active in MIMO-OSTBC environment. We show that the effective capacity decreases by the increase of interferers or their interference power. In addition, operation of MIMO-OSTBC with RAS was compared with MISO-OSTBC systems. We indicate that RAS scheme is more efficient in high QoS demand.
%%%%%%%%%%%%%%%%%%%%%%%%%%%%%%%%%%%%%%%%%%%%%%%%%
%%% REFERENCES %%%%%%%%%%%%%%%%%%%%%%%%%%%%%%%%%%
%%%%%%%%%%%%%%%%%%%%%%%%%%%%%%%%%%%%%%%%%%%%%%%%%

\end{document}